# Radiation pressure and the linear momentum of the electromagnetic field in magnetic media

Masud Mansuripur

*College of Optical Sciences, The University of Arizona, Tucson, Arizona 85721*
*masud@optics.arizona.edu*

**Abstract**: We examine the force of the electromagnetic radiation on linear, isotropic, homogeneous media specified in terms of their permittivity $\varepsilon$ and permeability $\mu$. A formula is proposed for the electromagnetic Lorentz force on the magnetization $\boldsymbol{M}$, which is treated here as an Amperian current loop. Using the proposed formula, we demonstrate conservation of momentum in several cases that are amenable to rigorous analysis based on the classical Maxwell equations, the Lorentz law of force, and the constitutive relations. Our analysis yields precise expressions for the density of the electro-magnetic and mechanical momenta inside the media that are specified by their $(\varepsilon,\mu)$ parameters. An interesting consequence of this analysis is the identification of an "intrinsic" mechanical momentum density, $\frac{1}{2}\boldsymbol{E}\times\boldsymbol{M}/c^2$, analogous to the electromagnetic (or Abraham) momentum density, $\frac{1}{2}\boldsymbol{E}\times\boldsymbol{H}/c^2$. (Here $E$ and $H$ are the magnitudes of the electric and magnetic fields, respectively, and $c$ is the speed of light in vacuum.) This intrinsic mechanical momentum, associated with the magnetization $\boldsymbol{M}$ in the presence of an electric field $\boldsymbol{E}$, is apparently the same "hidden" momentum that was predicted by W. Shockley and R. P. James nearly four decades ago.

**OCIS codes**: (260.2110) Electromagnetic theory; (140.7010) Trapping.

**References and links**

1. J. P. Gordon, "Radiation forces and momenta in dielectric media," Phys. Rev. A **8**, 14-21 (1973).
2. D. F. Nelson, "Momentum, pseudomomentum, and wave momentum: Toward resolving the Minkowski-Abraham controversy," Phys. Rev. A **44**, 3985 (1991).
3. I. Brevik, "Experiments in phenomenological electrodynamics and the electromagnetic energy-momentum. tensor," Physics Reports **52**, 133-201 (1979).
4. R. Loudon, "Theory of the radiation pressure on dielectric surfaces," J. Mod. Opt. **49**, 821-838 (2002).
5. R. Loudon, "Radiation pressure and momentum in dielectrics," Fortschr. Phys. **52**, 1134-1140 (2004).
6. R. Loudon, S. M. Barnett, and C. Baxter, "Radiation pressure and momentum transfer in dielectrics: the photon drag effect," Phys. Rev. A **71**, 063802 (2005).
7. M. Mansuripur, "Radiation pressure and the linear momentum of the electromagnetic field," Opt. Express **12**, 5375-5401 (2004).
8. M. Mansuripur, A. R. Zakharian, and J. V. Moloney, "Radiation pressure on a dielectric wedge," Opt. Express **13**, 2064-2074 (2005).
9. M. Mansuripur, "Radiation pressure and the linear momentum of light in dispersive dielectric media," Opt. Express **13**, 2245-2250 (2005).
10. M. Mansuripur, "Angular momentum of circularly polarized light in dielectric media," Opt. Express **13**, 5315-5324 (2005).
11. M. Mansuripur, "Radiation pressure and the distribution of electromagnetic force in dielectric media," SPIE Proc. **5930**, Optical Trapping and Optical Micromanipulation II, K. Dholakia and G. C. Spalding, Eds. (2005).
12. M. Mansuripur, A. R. Zakharian, and J. V. Moloney, "Equivalence of total force (and torque) for two formulations of the Lorentz law," *SPIE Proc.* **6326**, 63260G, Optical Trapping and Optical Micro-manipulation III, K. Dholakia and G. C. Spalding, Eds. (2006).
13. M. Mansuripur, "Radiation Pressure on Submerged Mirrors: Implications for the Momentum of Light in Dielectric Media," Opt. Express **15**, 2677-2682 (2007).
14. B. D. H. Tellegen, "Magnetic-Dipole Models," Am. J. Phys. **30**, 650 (1962).
15. S. M. Barnett and R. Loudon, "On the electromagnetic force on a dielectric medium," J. Phys. B: At. Mol. Opt. Phys. **39**, S671-S684 (2006).

16. B. Kemp, T. Grzegorczyk, and J. Kong, "Ab initio study of the radiation pressure on dielectric and magnetic media," Opt. Express **13**, 9280-9291 (2005).
17. B. A. Kemp, J. A. Kong, and T. Grzegorczyk, "Reversal of wave momentum in isotropic left-handed media," Phys. Rev. A **75**, 053810 (2007).
18. L. Vaidman, "Torque and force on a magnetic dipole," Am. J. Phys. **58**, 978-983 (1990).
19. A. D. Yaghjian, "Electromagnetic forces on point dipoles," IEEE Anten. Prop. Soc. Symp. **4**, 2868-2871 (1999).
20. W. Shockley, "Hidden linear momentum related to the $\alpha \cdot E$ term for a Dirac-electron wave packet in an electric field," Phys. Rev. Lett. **20**, 343-346 (1968).
21. M. Mansuripur, "Momentum of the electromagnetic field in transparent dielectric media," SPIE Proc. **6644**, Optical Trapping and Optical Micro-manipulation IV, K. Dholakia and G. C. Spalding, Eds. (2007).
22. W. Shockley and R. P. James, "Try simplest cases discovery of hidden momentum forces on magnetic currents," Phys. Rev. Lett. **18**, 876-879 (1967).
23. P. Penfield and H. A. Haus, "Electrodynamics of Moving Media," MIT Press, Cambridge (1967).
24. R. P. Feynman, R. B. Leighton, and M. Sands, "The Feynman Lectures on Physics," Vol. II, Chap. 27, Addison-Wesley, Reading, Massachusetts (1964).

**1. Introduction**

The magnitude of the momentum of light in dielectric media has been the subject of debate and controversy for the past hundred years. There have been several arguments, from theory and experiment, as to why the photon momentum inside a dielectric material should or should not be expressed by either of the two competing formulas associated with the names of H. Minkowski and M. Abraham [1-6]. In a series of papers published in recent years [7-13], we have argued that the correct expression for the photon momentum is neither Minkowski's nor Abraham's, but rather it is the arithmetic average of these two expressions. The present paper extends our argument to the case of linear, isotropic, homogeneous (LIH) magnetic materials, for which we derive expressions for the radiation pressure and momentum in terms of the material's permittivity $\varepsilon$ and permeability $\mu$.

Treating the magnetization density $M$ of a material medium as an Amperian current loop [14], we arrive in Sec. 2 at a specific expression for the force exerted by the electromagnetic field on $M$. Our belief in the validity of this expression stems from our analysis of radiation pressure on semi-infinite slabs, the results of which turn out to be in complete agreement with the momentum conservation law.

Computing the total electromagnetic force on a rigid body requires, in addition to the force exerted on $M$, the Lorentz force of the electromagnetic field on induced electrical charges and currents [4-7]. The force density on induced currents is $\boldsymbol{F} = (\partial \boldsymbol{P}/\partial t) \times \boldsymbol{B}$, where $\boldsymbol{P}$ is the polarization density of the medium, and $\boldsymbol{B} = \mu_o(\boldsymbol{H} + \boldsymbol{M})$ is the magnetic induction. As for the induced (bound) charge density $\rho_b = -\nabla \cdot \boldsymbol{P}$, the corresponding force density may be written as $\boldsymbol{F} = -(\nabla \cdot \boldsymbol{P})\boldsymbol{E}$. For the LIH media discussed in the present paper, $\nabla \cdot \boldsymbol{P} = \varepsilon_o(\varepsilon - 1)\nabla \cdot \boldsymbol{E}$ vanishes everywhere within the bulk of the medium, thus confining the force of the $E$-field to surfaces and interfaces (where the induced charge density can be non-zero). An alternative formula for the $E$-field's contribution to the Lorentz force density, $\boldsymbol{F} = (\boldsymbol{P} \cdot \nabla)\boldsymbol{E}$, has been used extensively in the literature [1-6]. The two formulations can be shown to yield the same *total* force (and torque) on rigid bodies [12,15], even though the force distribution obtained with $\boldsymbol{F} = -(\nabla \cdot \boldsymbol{P})\boldsymbol{E}$ can differ substantially from that obtained using $\boldsymbol{F} = (\boldsymbol{P} \cdot \nabla)\boldsymbol{E}$. In what follows, whenever the total force exerted by the $E$-field happens to be non-zero, we will present two sets of results, one for each formulation.

In Sec. 3, we use the complete expression of the Lorentz force on a semi-infinite LIH magnetic medium specified by $(\varepsilon, \mu)$, to determine the radiation pressure at the entrance facet as well as the momentum density of the light inside the medium. The "Einstein box" Gedanken experiment [5] described in Sec. 4 then sets the stage for the discussion in Sec. 5 of the nature of the optical momentum and its division into electromagnetic and mechanical parts.

In contrast to plane-waves, which are infinite in extent, a finite-diameter beam of light exerts a lateral force on its host medium at and around the lateral boundaries (i.e., sidewalls) of the beam [7]. The strength of this force depends on the polarization state of the beam as well as on the host material's $(\varepsilon, \mu)$ parameters; the direction of the force, which also depends on the aforementioned parameters, can be either expansive or compressive. In Sec. 6 we derive expressions for the optical force density at the sidewalls of finite-diameter beams inside a transparent medium (i.e., one for which both $\varepsilon$ and $\mu$ are real-valued and have the same sign).

Computing the force of a plane-wave at oblique incidence on a semi-infinite magnetic slab is the subject of Secs. 7 and 8, which address the cases of p- and s-polarized light. We derive formulas for the radiation pressure at the entrance facet of a semi-infinite slab, and show consistency with the results obtained in Sec. 3 for the case of normal incidence.

The great advantage of studying the momentum of plane-waves entering semi-infinite slabs from the free-space is that, in general, there are two alternative ways of calculating the radiation pressure, one based on the knowledge of the incident and reflected momenta in the free-space alone, the other based on the Lorentz force of the electromagnetic field that enters the dielectric/magnetic slab. Calculation of the force in the latter case, of course, is made possible by the specification of the slab's material in terms of idealized constitutive relations. The two methods of calculation must, in the end, yield identical results, or else the initial hypotheses concerning the nature of the elementary forces must be abandoned. The cases analyzed in the present paper are chosen for their essential simplicity, which enables one to obtain exact solutions of the Maxwell equations. The two methods are applied to each problem, and the consistency of the solutions in each and every case is demonstrated.

## 2. Lorentz force of the electromagnetic field on the magnetization of a medium

The magnetization $\boldsymbol{M}(x, y, z, t) = M_x \hat{\boldsymbol{x}} + M_y \hat{\boldsymbol{y}} + M_z \hat{\boldsymbol{z}}$ of a material at a given point in space and time is subject to the Lorentz force of the local magnetic field. The magnetic induction $\boldsymbol{B}$ exerts a force on electric currents, and since $\boldsymbol{M}$ is ultimately rooted in Amperian current loops on the atomic scale [14], it is natural to express the force of the $B$-field on $\boldsymbol{M}$ as the sum of contributions from all the various atomic currents that make up the $M$-field. A current $I$ circulating around a small loop of area $\delta^2$ produces a magnetic dipole moment $\boldsymbol{m} = I\delta^2 \hat{\boldsymbol{n}}$, where $\hat{\boldsymbol{n}}$ is a unit vector perpendicular to the loop's surface. Denoting the number density of the loops in the medium by $N$, we will have $\boldsymbol{M} = N\boldsymbol{m}$. Equivalently, one may assign a magnetic dipole moment $\boldsymbol{m} = \boldsymbol{M}\delta^3$ to each cubic region of volume $\delta^3$; the three loop currents depicted in Fig. 1 will then be $M_x\delta$, $M_y\delta$, and $M_z\delta$, respectively.

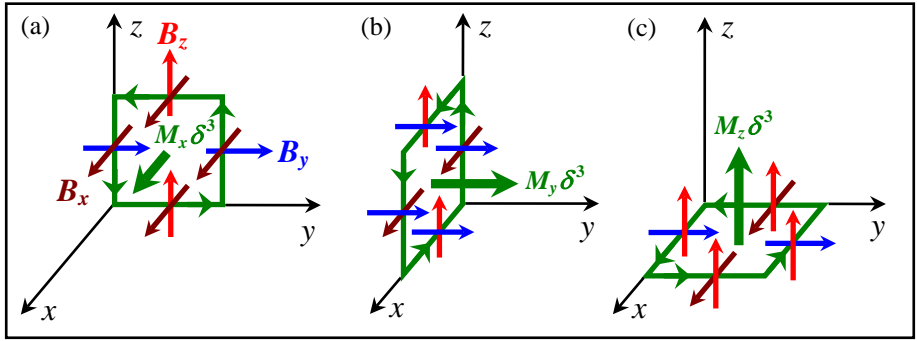

Fig. 1. The local magnetization $\boldsymbol{M} = M_x\hat{\boldsymbol{x}} + M_y\hat{\boldsymbol{y}} + M_z\hat{\boldsymbol{z}}$ of the material is subject to various local $B$-field components: $B_x$ (brown), $B_y$ (blue), and $B_z$ (red). A circulating current $I$ around a loop of area $\delta^2$ (green squares) produces a magnetic dipole $\boldsymbol{m} = I\delta^2 \hat{\boldsymbol{n}}$ along the perpendicular unit vector $\hat{\boldsymbol{n}}$. The magnetization density is $\boldsymbol{M} = N\boldsymbol{m}$, where $N$ is the number density of the loops.

Figure 1 shows three current loops representing the Cartesian components of $\boldsymbol{M}$ (green arrows) as well as the relevant components of $\boldsymbol{B}$ (brown, blue, and red arrows). According to the Lorentz law, the electromagnetic force on each leg of each loop is produced by the action of the local $B$-field. The various components of the force density (i.e., force per unit volume) for the loops of Fig. 1 may thus be written as follows:

$$\boldsymbol{F}_a = M_x(\partial B_x/\partial y)\hat{\boldsymbol{y}} + M_x(\partial B_x/\partial z)\hat{\boldsymbol{z}} - M_x(\partial B_y/\partial y)\hat{\boldsymbol{x}} - M_x(\partial B_z/\partial z)\hat{\boldsymbol{x}}, \tag{1a}$$

$$\boldsymbol{F}_b = M_y(\partial B_y/\partial x)\hat{\boldsymbol{x}} + M_y(\partial B_y/\partial z)\hat{\boldsymbol{z}} - M_y(\partial B_x/\partial x)\hat{\boldsymbol{y}} - M_y(\partial B_z/\partial z)\hat{\boldsymbol{y}}, \tag{1b}$$

$$\boldsymbol{F}_c = M_z(\partial B_z/\partial x)\hat{\boldsymbol{x}} + M_z(\partial B_z/\partial y)\hat{\boldsymbol{y}} - M_z(\partial B_x/\partial x)\hat{\boldsymbol{z}} - M_z(\partial B_y/\partial y)\hat{\boldsymbol{z}}. \tag{1c}$$

Adding the above forces together we find, after standard algebraic manipulations,

$$\boldsymbol{F}_m(x,y,z,t) = \boldsymbol{M}\times(\nabla\times\boldsymbol{B}) + (\boldsymbol{M}\cdot\nabla)\boldsymbol{B} - (\nabla\cdot\boldsymbol{B})\boldsymbol{M}. \tag{2}$$

The last term in Eq. (2) may be set to zero in accordance with Maxwell's equation $\nabla\cdot\boldsymbol{B}=0$. As for the remaining terms, we note that the defining relation $\boldsymbol{B}=\mu_o(\boldsymbol{H}+\boldsymbol{M})$ indicates that a certain fraction of the *B*-field is produced by the local magnetization $\boldsymbol{M}$. If we exclude this part of $\boldsymbol{B}$ from exerting a force on its own progenitor, we are left with $\mu_o\boldsol{H}$ as the effective field that exerts a force on the current loops. We thus have

$$\boldsymbol{F}_m(x,y,z,t) = \mu_o[\boldsymbol{M}\times(\nabla\times\boldsymbol{H}) + (\boldsymbol{M}\cdot\nabla)\boldsymbol{H}]. \tag{3a}$$

Equation (3a) is our basic formula under "steady-state" conditions (i.e., in the absence of transient events) for the Lorentz force density on the magnetization $\boldsymbol{M}$ of magnetic (or magnetizable) materials. When integrated over the volume of interest, Eq. (3a) should yield the total force exerted by the *H*-field on the magnetic dipoles of the material. In LIH materials where $\boldsymbol{M}=\chi\boldsymbol{H}$ and $\boldsymbol{B}=\mu_o(1+\chi)\boldsymbol{H}=\mu_o\mu\boldsymbol{H}$, the vector identity $\boldsymbol{A}\times(\nabla\times\boldsymbol{A})+(\boldsymbol{A}\cdot\nabla)\boldsymbol{A}=\tfrac{1}{2}\nabla(\boldsymbol{A}\cdot\boldsymbol{A})$ further simplifies Eq. (3a) as follows:

$$\boldsymbol{F}_m(x,y,z,t) = \tfrac{1}{2}\mu_o(\mu-1)\nabla(\boldsymbol{H}\cdot\boldsymbol{H}). \tag{3b}$$

In deriving Eq. (3a), no assumptions were made about $\boldsymbol{B}$, $\boldsymbol{H}$, and $\boldsymbol{M}$ beyond the defining relation $\boldsymbol{B}=\mu_o(\boldsymbol{H}+\boldsymbol{M})$ and the Maxwell equation $\nabla\cdot\boldsymbol{B}=0$. In this formulation there is no natural way to introduce the magnetic charge density, which is usually defined as $\rho_m=-\nabla\cdot\boldsymbol{M}$ and considered analogous to the bound electric charge density $\rho_b=-\nabla\cdot\boldsymbol{P}$ [16,17]. Whereas the electric charges (free or bound) are acted upon by the *E*-field in accordance with the Lorentz law $\boldsymbol{F}=\rho_b\boldsymbol{E}$, in our formulation there is no corresponding interaction between the magnetic charge density $\rho_m$ and the magnetic fields. Note, however, that when the field components whose derivatives appear in Eq. (3a) happen to be discontinuous at the boundaries and interfaces between adjacent media, one must be careful to account for the forces experienced by the magnetic dipoles at such boundaries.

The above expression for the Lorentz force on magnetization $\boldsymbol{M}$, combined with that pertaining to material polarization $\boldsymbol{P}$ discussed in Sec. 1, yields correct predictions for the radiation pressure in "steady-state" situations, as will be shown in Secs. 3, 6, 7 and 8 below. In problems that involve transient events, such as the passage of the leading or trailing edge of a light pulse through a magnetic medium (see Sec. 5), Eq. (3) must be augmented by an additional term, $\partial(\boldsymbol{E}\times\boldsymbol{M}/c^2)/\partial t$, to account for the "hidden" or "intrinsic" mechanical momentum produced by the action of the *E*-field on magnetic dipoles [18, 19]. While proper derivation of this additional force term requires a foray into quantum electrodynamics [20], the classical arguments presented in Secs. 4 and 5 provide ample justification for its existence. The addition of $\partial(\boldsymbol{E}\times\boldsymbol{M}/c^2)/\partial t$ to Eq. (3), of course, does not modify the final results of steady-state calculations, as the new term generally vanishes upon time-averaging.

The final equation that emerges from the above discussion of the electromagnetic force exerted on the magnetization $\boldsymbol{M}$ is analogous to that of the Lorentz force experienced by the polarization $\boldsymbol{P}$, namely, $\boldsymbol{F}_e=(\boldsymbol{P}\cdot\nabla)\boldsymbol{E}+(\partial\boldsymbol{P}/\partial t)\times\boldsymbol{B}$, which can be equivalently written as $\boldsymbol{F}_e=(\boldsymbol{P}\cdot\nabla)\boldsymbol{E}+\boldsymbol{P}\times(\nabla\times\boldsymbol{E})+\partial(\boldsymbol{P}\times\boldsymbol{B})/\partial t$. In LIH media, where $\boldsymbol{P}=\varepsilon_o(\varepsilon-1)\boldsymbol{E}$, the force density experienced by $\boldsymbol{P}$ will be $\boldsymbol{F}_e(x,y,z,t)=\tfrac{1}{2}\varepsilon_o(\varepsilon-1)\nabla(\boldsymbol{E}\cdot\boldsymbol{E})+\partial(\boldsymbol{P}\times\boldsymbol{B})/\partial t$.

## 3. Radiation pressure and momentum in magnetic media

Consider the stationary slab shown in Fig. 2, and assume that the slab has relative permittivity and permeability constants $\varepsilon$ and $\mu$, respectively. In general, $\varepsilon$ and $\mu$ are complex functions of the frequency $f$; only when transparent materials are considered shall we assume that both $\varepsilon$ and $\mu$ are real-valued. By convention, the real parts of $\varepsilon$ and $\mu$ could be positive or negative, but their imaginary parts are always greater than or equal to zero. In the following discussion both $\sqrt{\varepsilon\mu}$ and $\sqrt{\varepsilon/\mu}$ will appear in various expressions. Since $\sqrt{\varepsilon\mu}$ appears in a plane-wave's exponential phase-factor, one must always choose the root with a non-negative imaginary part, otherwise the wave-amplitude will increase indefinitely as $z \to \infty$. In particular, when both $\varepsilon$ and $\mu$ are real and negative (i.e., the case of "negative–index" materials), one may resort to a limiting argument to show that $\sqrt{\varepsilon\mu}$ must be a negative number. In contrast, the sign of $\sqrt{\varepsilon/\mu}$ is determined by requiring consistency among Maxwell's equations; in the case of negative-index materials, such considerations reveal the sign of $\sqrt{\varepsilon/\mu}$ as positive.

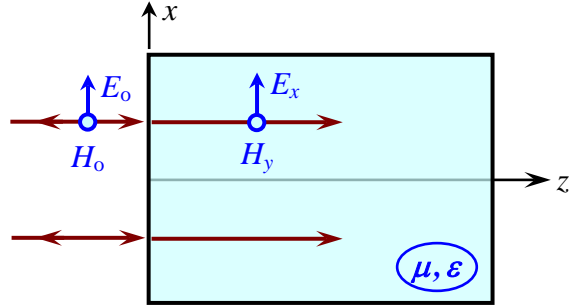

Fig. 2. A linearly polarized plane-wave having $E$-field amplitude $E_o$ and $H$-field amplitude $H_o$ is normally incident at the interface between the free space and a LIH slab of permittivity $\varepsilon$ and permeability $\mu$. The Fresnel reflection coefficient at the surface is $\rho$. Inside the slab, the transmitted beam has field amplitudes $E_x$ and $H_y$.

With reference to Fig. 2, a normally incident plane-wave, having complex phase-factor $\exp\{i2\pi f[(z/c) - t]\}$ and field amplitudes $E_o$ and $H_o = E_o/Z_o$, where $Z_o = \sqrt{\mu_o/\varepsilon_o}$, arrives at a semi-infinite slab of complex-valued $\varepsilon$ and $\mu$ from the free space region on the left-hand side. The Fresnel reflection coefficient at the entrance facet of the slab is readily found to be

$$\rho = (1 - \sqrt{\varepsilon/\mu})/(1 + \sqrt{\varepsilon/\mu}). \tag{4}$$

Inside the slab, the phase-factor is $\exp\{i2\pi f[\sqrt{\varepsilon\mu}(z/c) - t]\}$, while the field amplitudes are $E_x = (1+\rho)E_o$ and $H_y = (1+\rho)\sqrt{\varepsilon/\mu}E_o/Z_o$. Using the time-averaged Poynting vector component along $z$, namely, $<S_z> = \tfrac{1}{2}\mathrm{Re}(E_x H_y^*)$, it is easy to verify that the rate of flow of optical energy in the incident beam minus that in the reflected beam is exactly equal to the rate of flow of energy into the slab, namely,

$$<S_z> = \tfrac{1}{2}Z_o^{-1}|E_o|^2(1-|\rho|^2) = 2Z_o^{-1}|E_o|^2\mathrm{Re}(\sqrt{\varepsilon/\mu})/|1 + \sqrt{\varepsilon/\mu}|^2. \tag{5}$$

As for the rate of flow of optical momentum into the slab, one must account for the Lorentz force of the $B$-field on the bound current density $\boldsymbol{J}_b = \partial \boldsymbol{P}/\partial t$, as well as that on the magnetization $\boldsymbol{M} = (\mu - 1)\boldsymbol{H}$. According to Eq. (3a), the force density experienced by the magnetic dipoles of the slab in the system of Fig. 2 is $\mu_o \boldsymbol{M} \times (\nabla \times \boldsymbol{H})$; the second term vanishes for this geometry. [Note that the contribution of $\partial(\boldsymbol{E} \times \boldsymbol{M}/c^2)/\partial t$ may be ignored for now, as this term's time-average is zero under steady-state conditions. Also, the contribution of bound charges to the Lorentz force, $\rho_b \boldsymbol{E}$, is zero in the present example, as $\rho_b = 0$ both inside the slab and at its entrance facet.] The total force density is thus given by

$$\boldsymbol{F} = (\partial \boldsymbol{P}/\partial t) \times \boldsymbol{B} + \mu_o \boldsymbol{M} \times (\nabla \times \boldsymbol{H}). \tag{6}$$

The use of Maxwell's 2$^{nd}$ equation, $\nabla \times \boldsymbol{H} = \partial \boldsymbol{D}/\partial t$, simplifies the above expression to

$$F_z = \mu_o \varepsilon_o (\varepsilon - 1)(\partial E_x/\partial t)(\mu H_y) - \mu_o \varepsilon_o \varepsilon (\partial E_x/\partial t)[(\mu - 1)H_y]. \tag{7}$$

Using the complex notation for $E$- and $H$-fields, and assuming for the moment that $\varepsilon$ and $\mu$ are complex-valued, the time-averaged force density, integrated over the thickness of the slab (i.e., $z$ ranging from 0 to $\infty$), is evaluated as follows:

$$\begin{aligned}\langle F_z \rangle &= \tfrac{1}{2}\mu_o\varepsilon_o \mathrm{Re}[-\mathrm{i}2\pi f(\varepsilon-1)E_x\mu^* H_y^* + \mathrm{i}2\pi f \varepsilon E_x(\mu^*-1)H_y^*]\int_0^\infty \exp[-4\pi f(z/c)\mathrm{Im}\sqrt{\varepsilon\mu}]\,\mathrm{d}z \\ &= \tfrac{1}{4}\varepsilon_o|(1+\rho)E_o|^2 \mathrm{Im}[(\varepsilon-1)\mu^*\sqrt{\varepsilon^*/\mu^*} - \varepsilon(\mu^*-1)\sqrt{\varepsilon^*/\mu^*}]/\mathrm{Im}(\sqrt{\varepsilon\mu}) \\ &= \varepsilon_o|E_o|^2(1+|\varepsilon/\mu|)/|1+\sqrt{\varepsilon/\mu}|^2. \end{aligned} \tag{8}$$

Note that $F_z$ in Eq. (7) is force per unit volume, whereas $F_z$ in Eq. (8) represents force per unit surface area of the slab. The last line of Eq. (8) is precisely the result one would expect from a consideration of the rates of flow of incident and reflected momenta in the free space, namely,

$$\langle F_z \rangle = \tfrac{1}{2}\varepsilon_o|E_o|^2(1+|\rho|^2) = \varepsilon_o|E_o|^2(1+|\varepsilon/\mu|)/|1+\sqrt{\varepsilon/\mu}|^2. \tag{9}$$

The fact that Eqs. (8) and (9) yield identical expressions for the radiation pressure on a slab specified by $\varepsilon$ and $\mu$, is an extremely important factor in support of the Lorentz force formula of Eq. (6) and, by extension, of Eq. (3). Similar equalities obtained in Secs. 7 and 8, in cases of oblique incidence with p- and s-polarized light, lend further credence to Eq. (3) as the correct expression for the Lorentz force exerted by the electromagnetic field on the material's magnetization $\boldsymbol{M}$ under steady-state conditions.

Although Eq. (8) was derived for the case of complex $\varepsilon$ and $\mu$, it turns out to be valid even when these parameters are real-valued; the reason being that the term $\mathrm{Im}(\sqrt{\varepsilon\mu})$ appearing (upon integration over $z$) in the denominator in the second line of Eq. (8), cancels out in the end; thus the fact that $\mathrm{Im}(\sqrt{\varepsilon\mu}) \to 0$ for a transparent medium does not affect the final result. In a transparent medium where $\varepsilon$ and $\mu$ are real-valued (both positive or both negative), $\langle F_z \rangle$ of Eq. (8) should be interpreted as the time-averaged rate of momentum flow at any cross-section of the beam inside the slab. In terms of the field amplitudes $|E_x|$ and $|H_y|$ within the transparent medium, Eq. (8) may be rewritten as

$$\langle F_z \rangle = \tfrac{1}{4}\varepsilon_o Z_o|E_x||H_y|\sqrt{\mu/\varepsilon}\,[1+(\varepsilon/\mu)]. \tag{10}$$

The momentum flux inside a transparent medium given by Eq. (10) is generally positive, as both $\varepsilon/\mu$ and $\sqrt{\mu/\varepsilon}$ are positive entities, whether $\varepsilon$ and $\mu$ are both positive or both negative. In a transparent, dispersionless medium where the speed of light is $V = c/\sqrt{\varepsilon\mu}$ (in this case $\varepsilon$ and $\mu$ are necessarily positive, as negative-index materials cannot be free from dispersion), the momentum per unit volume becomes

$$\boldsymbol{p} = \langle F_z \rangle \hat{\boldsymbol{z}}/V = \tfrac{1}{4}(\mu+\varepsilon)|E_x||H_y|\hat{\boldsymbol{z}}/c^2 = \tfrac{1}{4}\varepsilon_o \boldsymbol{E}\times\boldsymbol{B} + \tfrac{1}{4}\mu_o \boldsymbol{D}\times\boldsymbol{H}. \tag{11}$$

This is precisely the result one would expect from the "gap" argument of Ref. [13]. In the case of non-magnetic dielectrics, where $\mu = 1$, the two terms on the right-hand side of Eq. (11) correspond, respectively, to the Abraham and Minkowski momentum densities, yielding a net density that is half-way between the two, as has been argued in our previous papers [7,13]. In the case of magnetic materials, however, the correspondence with Abraham and Minkowski momenta breaks down, and the correct formulation is simply that given by Eq. (11).

With real-valued $\varepsilon$ and $\mu$, Eqs. (5) and (9) yield the rates of flow of optical energy and momentum into a transparent slab. For an incident pulse of duration $\tau$ and cross-sectional area $A$, the number of photons entering the slab will be $<S_z>A\tau/(hf)$, where $h$ is Planck's constant; therefore, the momentum of each photon inside the (transparent) material will be

$$p_{\text{photon}} = \frac{<F_z>A\tau}{<S_z>A\tau/(hf)} = \tfrac{1}{2}(\sqrt{\varepsilon/\mu} + \sqrt{\mu/\varepsilon})(hf/c). \quad (12)$$

The above formula for photon momentum inside a transparent medium of relative permittivity $\varepsilon$ and permeability $\mu$ is applicable to both positive- and negative-index media (in both cases $\varepsilon/\mu > 0$). The value of $p_{\text{photon}}$ in Eq. (12) is always greater than or equal to the free-space momentum $hf/c$. For a proof, note that $\tfrac{1}{2}(\sqrt{\varepsilon/\mu} + \sqrt{\mu/\varepsilon}) \geq 1$ leads to $1 + (\varepsilon/\mu) \geq 2\sqrt{\varepsilon/\mu}$, which is equivalent to $(1 - \sqrt{\varepsilon/\mu})^2 \geq 0$, an obviously valid inequality.

The entire argument of Ref. [21] with regard to a pulse of light entering and exiting a transparent, dispersionless slab may now be repeated with a material having arbitrary $\varepsilon$ and $\mu$ (both real and positive, of course, as negative-index materials cannot be free from dispersion), without changing the final conclusions. In Ref. [21] we also consider the case of a light pulse entering an antireflection-coated (AR) slab, and show that the excess photon momentum inside the (transparent) slab is consistent with the negative force exerted on the AR layer during the time in which the pulse enters the slab. A similar argument can now be made for transparent magnetic media; here the AR coating must be a quarter-wave-thick layer of a transparent material having permittivity $\sqrt{|\varepsilon|}$ and permeability $\sqrt{|\mu|}$. A straightforward calculation similar to that in Ref. [7] for dielectric media now yields the following expression for the force per unit area of the AR-coating layer:

$$<F_z> = \tfrac{1}{4}\varepsilon_o|E_o|^2(1 - \sqrt{\varepsilon/\mu})(1 - \sqrt{\mu/\varepsilon}). \quad (13)$$

The above force is always negative (i.e., the incident light from the free space pulls on the AR coating layer), because either $\varepsilon/\mu > 1$ or $\mu/\varepsilon > 1$. (When $\varepsilon = \mu$, the slab's Fresnel reflection coefficient is zero and, therefore, there is no need for AR coating.) The negative force on the AR coating layer given by Eq. (13) accounts for the excess photon momentum when it enters from the free space into a transparent slab.

## 4. Division of momentum into electromagnetic and mechanical parts

A light pulse traveling inside a transparent medium has both electromagnetic and mechanical momenta. The mechanical momentum is due to the motion of the atoms/molecules of the medium in response to the electromagnetic forces exerted upon them by the light pulse. In a variant of the "Einstein box" Gedanken experiment [5] depicted in Fig. 3, a pulse of energy $E = mc^2$ and (free-space) momentum $\boldsymbol{p} = mc\hat{z}$ travels either outside or inside a transparent slab of length $L$ and mass $M_o$. The entrance and exit facets of the slab are anti-reflection coated to ensure the passage of the entire pulse through the slab, with no reflection losses whatsoever. The pulse crosses the slab in a time interval $\Delta t = L/V_g$, where the group velocity $V_g$ is a function of the optical frequency $f$ and the (frequency-dependent) material parameters $\varepsilon$ and $\mu$.

When the pulse travels outside and the slab is stationary, the center of mass of the system moves along the $z$-axis at the constant velocity $V_{CM} = mc/(m + M_o)$. The displacement of the center of mass during a time interval $\tau$ is, therefore, $mc\tau/(m + M_o)$. If the pulse goes through the slab, however, its velocity, while inside the slab, will drop down to the group velocity $V_g$. The emergent pulse will thus stay behind the pulse that has traveled in the free-space by a distance $[(c/V_g) - 1]L$, as shown in Fig. 3. Therefore, for the system's center of mass to be in the same place in both experiments, it is necessary for the slab in the latter case to have shifted to the right by $\Delta z = [(c/V_g) - 1]Lm/M_o$. This displacement, which occurs during the time interval $\Delta t = L/V_g$ when the pulse is inside the slab, requires the slab's mechanical momentum during the passage of the pulse to be

$$\boldsymbol{p}_M = M_o(\Delta z/\Delta t)\hat{z} = m(c - V_g)\hat{z}. \quad (14)$$

Since the total momentum of the system is that of the pulse before entering the slab, namely, $\boldsymbol{p} = mc\hat{z}$, we conclude that the pulse's electromagnetic momentum inside the slab must be

$$\boldsymbol{p}_E = mV_g\hat{z} = (E/c)(V_g/c)\hat{z}. \tag{15}$$

In other words, the pulse's electromagnetic momentum within the medium is reduced by a factor of $V_g/c$ relative to its free-space value [5,9]. In a dispersionless medium where $V_g = c/n$, $n$ being the refractive index, the electromagnetic momentum of the pulse will be $\boldsymbol{p}_E = (E/nc)\hat{z}$, which is commonly referred to as the Abraham momentum. The difference between the free-space momentum of the pulse and its electromagnetic (or Abraham) momentum is thus transferred to the slab in the form of mechanical momentum, $\boldsymbol{p}_M$, causing the slab's eventual displacement in a manner consistent with the demands of the Einstein box experiment.

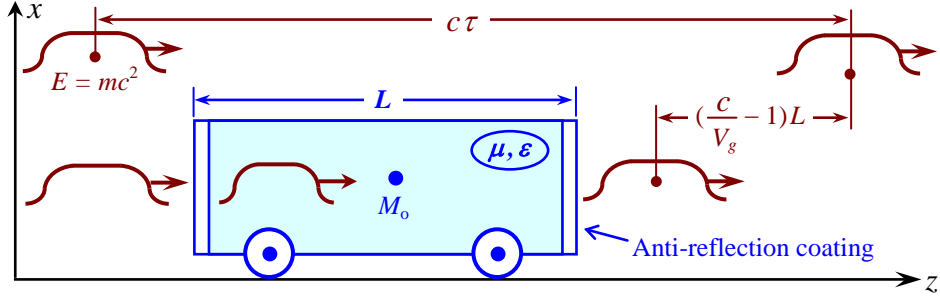

Fig. 3. Variant of the Einstein box experiment featuring a short pulse of light and a transparent slab of length $L$ and mass $M_o$. In the free-space region outside the slab, the pulse, having energy $E = mc^2$ and momentum $\boldsymbol{p} = mc\hat{z}$, travels with speed $c$. Inside the slab, the pulse travels with the group velocity $V_g$. The entrance and exit facets of the slab are anti-reflection coated to ensure the passage of the entire pulse through the slab. In one experiment, the pulse travels entirely in the free-space region outside the slab, while in another, the pulse spends a fraction of its time inside the slab. Since no external forces are at work, the center of mass of the system (consisting of the light pulse and the slab) must be displaced equally in the two experiments.

## 5. Mechanical momentum in dispersionless magnetic media

We derive the mechanical momentum density inside a transparent, dispersionless, magnetic material using two different methods. The two methods yield answers that differ by $\tfrac{1}{2}\boldsymbol{E}\times\boldsymbol{M}/c^2$, thus suggesting the existence of an *intrinsic* mechanical momentum associated with the magnetization of the medium. As it turns out, this seemingly implausible momentum was predicted nearly forty years ago by Shockley and James [22] and, independently, by Penfield and Haus [23]. Our inference of the intrinsic mechanical momentum density, $\tfrac{1}{2}\boldsymbol{E}\times\boldsymbol{M}/c^2$, is based on purely classical arguments, relying on Maxwell's equations, standard constitutive relations, the expression of the Lorentz force on $\boldsymbol{M}$ given in Eq. (3), and the conclusions reached from the "Einstein box" Gedanken experiment of Sec. 4. The relevant physics in which the inherent momentum is rooted, however, is quantum electrodynamics, as argued by Shockley [20], and as will be discussed briefly at the end of the present section.

Our first method of calculating the mechanical momentum of a light pulse inside a transparent, dispersionless, magnetic medium considers the forces exerted on the medium by the leading and trailing edges of the pulse. (The present discussion is limited to media for which $\varepsilon$ and $\mu$ are real and positive, as lack of dispersion excludes negative-index media from such considerations. However, a treatment similar to that of Ref. [9] shows that our essential conclusions remain valid for dispersive media as well, including those that have a negative index.) In a transparent, dispersionless medium, Eq. (7) may be written as follows:

$$F_z(z,t) = \mu_o\varepsilon_o[(\varepsilon-1)\mu - (\mu-1)\varepsilon]H_y\partial E_x/\partial t$$
$$= \mu_o\varepsilon_o(\varepsilon-\mu)Z_o^{-1}\sqrt{\varepsilon/\mu}\,E_x\partial E_x/\partial t$$

$$= \varepsilon_\text{o}[1-(\varepsilon/\mu)]E_x \partial E_x/\partial z$$
$$= \tfrac{1}{2}\varepsilon_\text{o}[1-(\varepsilon/\mu)]\partial E_x^2/\partial z. \tag{16}$$

The identities $H_y = Z_\text{o}^{-1}\sqrt{\varepsilon/\mu}\,E_x$ and $\partial E_x/\partial t + (c/\sqrt{\varepsilon\mu})\partial E_x/\partial z = 0$ have been used in the above derivation, the latter being a direct consequence of dispersionless propagation along $z$ at the constant velocity $c/\sqrt{\varepsilon\mu}$. Integrating the force density of Eq. (16) along the $z$-axis, from the mid-point of the pulse, say, $z = z_\text{o}$ to $z = +\infty$ (leading edge) or $z = -\infty$ (trailing edge) yields

$$F_z(t) = \pm\tfrac{1}{2}\varepsilon_\text{o}[(\varepsilon/\mu)-1]E_x^2(z=z_\text{o},t). \tag{17}$$

Thus the time-averaged force per unit cross-sectional area, exerted on the medium by these edges of the pulse will be

$$\langle F_z \rangle = \pm\tfrac{1}{4}\varepsilon_\text{o}[(\varepsilon/\mu)-1]\,|E_x|^2. \tag{18}$$

Here the + and − signs apply to the leading and trailing edges, respectively. Equation (18) clearly indicates the equal but opposite nature of the forces exerted on the medium by the pulse's front and rear edges. Ignoring the acoustic propagation of the pressure wave thus generated at the edges of the pulse, the mechanical momentum imparted to the medium is seen to reside between the front and rear edges. Dividing the force density of Eq. (18) by the speed of light $c/\sqrt{\varepsilon\mu}$ in the dispersionless medium yields the volume density of the mechanical momentum (imparted to the medium by the leading edge) as follows:

$$\boldsymbol{p}_{mech} = \tfrac{1}{4}\mu_\text{o}\varepsilon_\text{o}(\varepsilon-\mu)|E_x||H_y|\hat{\boldsymbol{z}} = \tfrac{1}{4}(\varepsilon-\mu)\boldsymbol{E}\times\boldsymbol{H}/c^2. \tag{19}$$

Our second method of calculating the mechanical momentum starts with Eq. (11) and proceeds by substituting $\mu_\text{o}(\boldsymbol{H}+\boldsymbol{M})$ for $\boldsymbol{B}$ and $(\varepsilon_\text{o}\boldsymbol{E}+\boldsymbol{P})$ for $\boldsymbol{D}$. The momentum density of a plane-wave inside a transparent, dispersionless medium may thus be written in the following equivalent form:

$$\boldsymbol{p} = \tfrac{1}{2}(\boldsymbol{E}\times\boldsymbol{H}/c^2) + \tfrac{1}{4}(\boldsymbol{E}\times\boldsymbol{M}/c^2) + \tfrac{1}{4}(\mu_\text{o}\boldsymbol{P}\times\boldsymbol{H}). \tag{20}$$

The first term on the right-hand side of Eq. (20) is the electromagnetic (or Abraham) momentum density in a non-dispersive material. The combined second and third terms, therefore, must provide the mechanical momentum density, namely,

$$\boldsymbol{p}_\text{mech} = \tfrac{1}{4}(\varepsilon+\mu-2)\boldsymbol{E}\times\boldsymbol{H}/c^2. \tag{21}$$

Equation (21) is a direct consequence of Eq. (11) in conjunction with the constraint imposed by the Einstein box experiment of Sec. 4. The mechanical momentum density of Eq. (21) is seen to be greater than that in Eq. (19) by $\tfrac{1}{2}\boldsymbol{E}\times\boldsymbol{M}/c^2$, which appears to be some sort of mechanical momentum inherent in the magnetization of the material. In other words, just as the electromagnetic (or Abraham) momentum density inside the medium is $\tfrac{1}{2}\boldsymbol{E}\times\boldsymbol{H}/c^2$, there appears to reside within the medium an "intrinsic" mechanical momentum density of $\tfrac{1}{2}\boldsymbol{E}\times\boldsymbol{M}/c^2$ as well. It is the sum of this intrinsic mechanical momentum and the momentum imparted to the medium by the leading edge of the pulse, given by Eq. (19), that produces the total mechanical momentum of Eq. (21). It thus appears that the missing force density at the pulse edges must be $F_z(z,t) = \partial(E_x M_y/c^2)/\partial t$, which was mentioned toward the end of Sec. 2. When this new term is added to Eq. (16), the resulting mechanical momentum density in Eq. (19) will coincide with that given by Eq. (21). We emphasize that this additional force density does not modify the steady-state analysis of Sec. 3, as the time-averaged value of the new term would be zero everywhere. For the same reason, the new term will be ignored in the sections that follow, as the discussion in these sections is confined to steady-state situations.

Shockley and James, based on entirely different arguments, have conjectured the existence of a similar, intrinsic mechanical momentum in magnetic media [22]. Shockley also provided an argument based on Dirac's theory of quantum electrodynamics in support of what he refers to as "hidden" momentum [20]. The essence of the argument is tied to Poynting's theorem, revolving around the fact that an electric field cannot exchange energy with an Amperian current loop that constitutes a magnetic dipole moment. (This, by the way, is the reason why the correct expression for the Poynting vector is $\boldsymbol{E}\times\boldsymbol{H}$ rather than $\boldsymbol{E}\times\boldsymbol{B}/\mu_\text{o}$ [14]. Several authors, the incomparable Richard Feynman among them [24], have failed to appreciate this subtle point, and consequently have arrived at the erroneous form of the Poynting vector.) The energy given by the local $E$-field to one leg of an Amperian current loop (constituting a magnetic dipole) must somehow be transferred to the opposite leg of the same loop (i.e., the leg whose current flows in the opposite direction). This internal exchange of energy imparts a net mechanical momentum to the loop, thus accounting for the "hidden" or "intrinsic" momentum density of $\tfrac{1}{2}\boldsymbol{E}\times\boldsymbol{M}/c^2$, which turned up in the above discussion.

We have shown in the present section that a classical treatment of radiation pressure in magnetic media (in conjunction with the Einstein box argument) demands the existence of an intrinsic $\tfrac{1}{2}\boldsymbol{E}\times\boldsymbol{M}/c^2$ mechanical momentum density. This momentum density in turn requires the addition of a new term, $\partial(\boldsymbol{E}\times\boldsymbol{M}/c^2)/\partial t$, to the classical formula for the Lorentz force density on magnetization $\boldsymbol{M}$, namely, Eq. (3). The physical mechanism responsible for the additional force is believed to be some sort of energy flux inside the Amperian current loops of magnetic dipoles, a quantum mechanical phenomenon that lies outside the domain of classical electrodynamics.

### 6. Lateral pressure at the sidewalls of a finite-diameter beam

Consider a collimated beam having a large but finite width along the $x$-axis, as shown in Fig. 4. When the beam is linearly polarized with its $H$-field along the $y$-axis (i.e., the case of transverse magnetic, TM, or p-polarization), the force density may be written as

$$\begin{aligned}\boldsymbol{F} &= (\partial\boldsymbol{P}/\partial t)\times\boldsymbol{B} + \mu_\text{o}\boldsymbol{M}\times(\nabla\times\boldsymbol{H}) \\ &= (\varepsilon-1)\varepsilon^{-1}(\partial\boldsymbol{D}/\partial t)\times(\mu_\text{o}\mu\boldsymbol{H}) + \mu_\text{o}(\mu-1)\boldsymbol{H}\times(\nabla\times\boldsymbol{H}) \\ &= \mu_\text{o}[1-(\mu/\varepsilon)]\,[(\partial H_y/\partial x)\hat{\boldsymbol{z}} - (\partial H_y/\partial z)\hat{\boldsymbol{x}}]\times H_y\hat{\boldsymbol{y}} \\ &= \tfrac{1}{2}\mu_\text{o}[(\mu/\varepsilon)-1]\,[(\partial H_y^2/\partial x)\hat{\boldsymbol{x}} + (\partial H_y^2/\partial z)\hat{\boldsymbol{z}}].\end{aligned} \quad (22)$$

Here Maxwell's 2$^\text{nd}$ equation, $\nabla\times\boldsymbol{H} = \partial\boldsymbol{D}/\partial t$, has been used in going from the second to the third line. Integrating the lateral force component $F_x$ from $x = 0$ to $\infty$ yields

$$\langle\boldsymbol{F}_{sw}^{(p)}\rangle = \int_0^\infty F_x\,\text{d}x\,\hat{\boldsymbol{x}} = \tfrac{1}{2}\mu_\text{o}[(\mu/\varepsilon)-1]\int_0^\infty(\partial H_y^2/\partial x)\,\text{d}x\,\hat{\boldsymbol{x}} = \tfrac{1}{2}\mu_\text{o}[1-(\mu/\varepsilon)]H_y^2(x=0,y,z,t)\hat{\boldsymbol{x}}. \quad (23)$$

Time-averaging the value of $H_y^2$ at the beam center, and using the relation $|E_x| = Z_\text{o}\sqrt{\mu/\varepsilon}\,|H_y|$, we obtain

$$\langle\boldsymbol{F}_{sw}^{(p)}\rangle = \pm\tfrac{1}{4}\mu_\text{o}[1-(\mu/\varepsilon)]\,|H_y|^2\,\hat{\boldsymbol{x}} = \pm\tfrac{1}{4}\varepsilon_\text{o}[(\varepsilon/\mu)-1]\,|E_x|^2\,\hat{\boldsymbol{x}}. \quad (24)$$

The ± signs in the above equation apply to the upper and lower sidewalls, respectively. The lateral force of Eq. (24) is expansive when $\varepsilon > \mu$, and compressive otherwise. Upon setting $\mu = 1$, the above formula reduces to the result obtained for non-magnetic dielectrics in Ref. [7].

An alternative formula for the Lorentz force on bound electric charges uses the term $(\boldsymbol{P}\cdot\nabla)\boldsymbol{E}$ in addition to those already present in Eq. (22). The additional force density will be

$$\boldsymbol{F}^{(\text{extra})} = (\boldsymbol{P}\cdot\nabla)\boldsymbol{E} = \varepsilon_\text{o}(\varepsilon-1)[(E_x\partial E_x/\partial x + E_z\partial E_x/\partial z)\hat{\boldsymbol{x}} + (E_x\partial E_z/\partial x + E_z\partial E_z/\partial z)\hat{\boldsymbol{z}}]. \quad (25)$$

From Maxwell's 2$^\text{nd}$ and 3$^\text{rd}$ equations we have $\partial E_x/\partial z = \partial E_z/\partial x - \mu_\text{o}\mu\partial H_y/\partial t$ and $\partial H_y/\partial x = \varepsilon_\text{o}\varepsilon\partial E_z/\partial t$. We also know that $\partial(E_zH_y)/\partial t = E_z\partial H_y/\partial t + H_y\partial E_z/\partial t$ and that $E_zH_y = -S_x$, where $S_x$ is the $x$-component of the Poynting vector. Substitution in Eq. (25) then yields

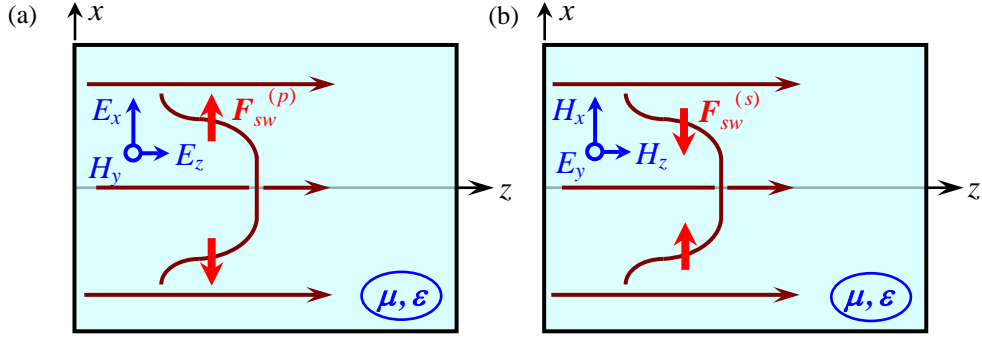

Fig. 4. Collimated beam of light propagating along $z$ and having a finite diameter along $x$. In (a) the beam is p-polarized, that is, its field components are $(E_x, E_z, H_y)$. In (b) the polarization state is s, corresponding to the field components $(E_y, H_x, H_z)$. The lateral electromagnetic force at the beam's sidewalls, denoted by $\boldsymbol{F}_{sw}$, is oriented in opposite directions on opposite sidewalls.

$$\begin{aligned}F_x^{(\text{extra})} &= \varepsilon_0(\varepsilon-1)(E_x \partial E_x/\partial x + E_z \partial E_z/\partial x - \mu_0\mu E_z \partial H_y/\partial t) \\ &= \varepsilon_0(\varepsilon-1)[\tfrac{1}{2}\partial E_x^2/\partial x + \tfrac{1}{2}\partial E_z^2/\partial x - \mu_0\mu\, \partial(E_z H_y)/\partial t + \mu_0\mu H_y \partial E_z/\partial t] \\ &= \varepsilon_0(\varepsilon-1)[\tfrac{1}{2}\partial(E_x^2+E_z^2)/\partial x + \mu_0\mu(\partial S_x/\partial t) + \tfrac{1}{2}(\mu_0/\varepsilon_0)(\mu/\varepsilon)\partial H_y^2/\partial x]. \end{aligned} \quad (26)$$

Integrating over $x$ from 0 to $\infty$, averaging over time, and setting $E_z(x=0,y,z,t)=0$, $\langle E_x^2(x=0,y,z,t)\rangle = \tfrac{1}{2}|E_x|^2$, and $\langle H_y^2(x=0,y,z,t)\rangle = \tfrac{1}{2}|H_y|^2$, we find

$$\left\langle \int_0^\infty F_x^{(\text{extra})}\, dx \right\rangle = -\tfrac{1}{2}\varepsilon_0(\varepsilon-1)|E_x|^2. \quad (27)$$

Finally, adding the above term to Eq. (24) yields the force on the beam's sidewalls as

$$\langle \boldsymbol{F}_{sw}^{(p)}\rangle = \pm\tfrac{1}{4}\varepsilon_0[(\varepsilon/\mu)-2\varepsilon+1]|E_x|^2 \hat{\boldsymbol{x}}. \quad (28)$$

Depending on whether the coefficient $[(\varepsilon/\mu)-2\varepsilon+1]$ in Eq. (28) is positive or negative, this sidewall force will be expansive or compressive. Equations (24) and (28) provide alternative expressions for radiation pressure at the sidewalls of a p-polarized beam. The former applies when the $E$-field contribution to the Lorentz force is expressed as $\boldsymbol{F} = -(\nabla \cdot \boldsymbol{P})\boldsymbol{E}$; the latter when it is $(\boldsymbol{P}\cdot\nabla)\boldsymbol{E}$. Both expressions apply to negative-index as well as positive-index media.

Calculation of the lateral pressure on the beam's sidewalls for s-polarized light (also known as transverse electric, TE) proceeds along similar lines. Here the field components are $(E_y, H_x, H_z)$, and the force density in accordance with Eq. (3a) is written

$$\begin{aligned}\boldsymbol{F} &= (\partial \boldsymbol{P}/\partial t)\times \boldsymbol{B} + \mu_0 \boldsymbol{M}\times(\nabla\times\boldsymbol{H}) \\ &\quad + \mu_0[(M_x \partial H_x/\partial x + M_z \partial H_x/\partial z)\hat{\boldsymbol{x}} + (M_x \partial H_z/\partial x + M_z \partial H_z/\partial z)\hat{\boldsymbol{z}}]. \end{aligned} \quad (29)$$

Using Maxwell's relations among the various field components, namely, $\partial E_y/\partial x = -\partial B_z/\partial t$, $\partial E_y/\partial z = \partial B_x/\partial t$, and $\partial H_x/\partial z - \partial H_z/\partial x = \partial D_y/\partial t$, the $x$-component of the force density in Eq. (29) is found to be

$$F_x = \varepsilon_0\mu_0\mu(\varepsilon-1)\partial(E_y H_z)/\partial t + \tfrac{1}{2}\varepsilon_0(\varepsilon-1)\partial E_y^2/\partial x + \tfrac{1}{2}\mu_0(\mu-1)\partial(H_x^2+H_z^2)/\partial x. \quad (30)$$

Integrating the above expression from $x=0$ to $\infty$, recognizing that the $x$-component of the Poynting vector, $S_x = E_y H_z$, time-averages to zero (i.e., no net lateral energy flux), and using the fact that at the beam center $H_z(x=0,y,z,t)$ vanishes, while $\langle E_y^2(x=0,y,z,t)\rangle = \tfrac{1}{2}|E_y|^2$, and $\langle H_x^2(x=0,y,z,t)\rangle = \tfrac{1}{2}|H_x|^2 = \tfrac{1}{2}(\varepsilon_0/\mu_0)(\varepsilon/\mu)|E_y|^2$, we find

$$\langle \boldsymbol{F}_{sw}^{(s)}\rangle = \pm\tfrac{1}{4}\varepsilon_0[(\varepsilon/\mu)-2\varepsilon+1]|E_y|^2 \hat{\boldsymbol{x}}. \quad (31)$$

In the above formula, which applies to both positive- and negative-index media, the + and − signs correspond to the upper and lower sidewalls, respectively. The lateral pressure at the sidewalls of an s-polarized beam thus turns out to be the same as that for a p-polarized beam given by Eq. (28). Whereas in the case of p-light the final result depends on whether the $E$-field contribution to the force is expressed as $-(\nabla \cdot \boldsymbol{P})\boldsymbol{E}$ or $(\boldsymbol{P} \cdot \nabla)\boldsymbol{E}$, for s-light the lateral pressure given by Eq. (31) does not distinguish between the two alternatives, as the $E$-field in the latter case makes no contribution to the force.

## 7. Oblique incidence on a magnetic slab – case of p-polarization

The case of oblique incidence on a magnetic slab provides further evidence for the validity of the Lorentz force expression on magnetic dipole moments, Eq. (3). First, we consider the case of a p-polarized plane-wave at oblique incidence on the interface between the free space and a LIH medium having permittivity $\varepsilon$ and permeability $\mu$, as shown in Fig. 5(a). The various field components are listed below, with the subscripts $i$, $r$, $t$ referring to the incident, reflected, and transmitted beams:

$$\boldsymbol{E}_i = E_o(\cos\theta\,\hat{\boldsymbol{x}} - \sin\theta\,\hat{\boldsymbol{z}})\exp\{i2\pi f[(x\sin\theta + z\cos\theta)/c - t]\} \tag{32a}$$

$$\boldsymbol{H}_i = Z_o^{-1}E_o\hat{\boldsymbol{y}}\exp\{i2\pi f[(x\sin\theta + z\cos\theta)/c - t]\} \tag{32b}$$

$$\boldsymbol{E}_r = \rho_p E_o(\cos\theta\,\hat{\boldsymbol{x}} + \sin\theta\,\hat{\boldsymbol{z}})\exp\{i2\pi f[(x\sin\theta - z\cos\theta)/c - t]\} \tag{32c}$$

$$\boldsymbol{H}_r = -Z_o^{-1}\rho_p E_o\hat{\boldsymbol{y}}\exp\{i2\pi f[(x\sin\theta - z\cos\theta)/c - t]\} \tag{32d}$$

$$\boldsymbol{E}_t = E_o[(1+\rho_p)\cos\theta\,\hat{\boldsymbol{x}} - \varepsilon^{-1}(1-\rho_p)\sin\theta\,\hat{\boldsymbol{z}}]\exp\{i2\pi f[(x\sin\theta + z\sqrt{\mu\varepsilon - \sin^2\theta})/c - t]\} \tag{32e}$$

$$\boldsymbol{H}_t = Z_o^{-1}(1-\rho_p)E_o\hat{\boldsymbol{y}}\exp\{i2\pi f[(x\sin\theta + z\sqrt{\mu\varepsilon - \sin^2\theta})/c - t]\}. \tag{32f}$$

In the above equations, the Fresnel reflection coefficient $\rho_p$ for a p-polarized plane-wave is

$$\rho_p = (\sqrt{\mu\varepsilon - \sin^2\theta} - \varepsilon\cos\theta)/(\sqrt{\mu\varepsilon - \sin^2\theta} + \varepsilon\cos\theta). \tag{33}$$

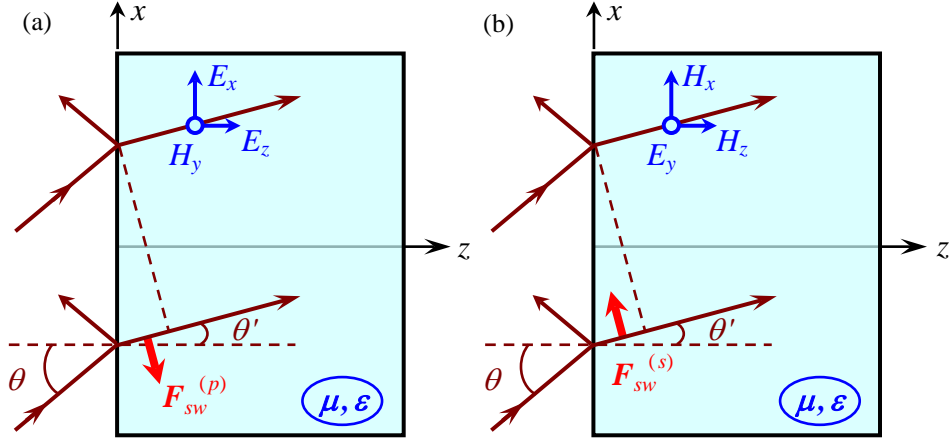

Fig. 5. Linearly polarized plane-wave incident at oblique angle $\theta$ at the interface between the free space and a homogeneous slab of permittivity $\varepsilon$ and permeability $\mu$. The incident beam is p-polarized in (a) and s-polarized in (b). In both cases the foot-print of the beam along $x$ is assumed to be unity, making the cross-sections of the incident and transmitted beams equal to $\cos\theta$ and $\cos\theta'$, respectively. The transmitted beam's lower sidewall has a segment of length $\sin\theta'$, which is subject to the sidewall force density $\boldsymbol{F}_{sw}$.

The force exerted on the bulk of the material is obtained by integrating Eq. (6) over the (infinite) thickness of the slab. There is also a surface force due to the bound charges at the interface of the slab with the free space [7]. After standard manipulations we find

$$F^{(\text{bulk})} = \frac{\varepsilon_o|E_o|^2\cos^2\theta}{|\sqrt{\mu\varepsilon-\sin^2\theta}+\varepsilon\cos\theta|^2}\{2\operatorname{Re}[\sqrt{\mu\varepsilon-\sin^2\theta}]\sin\theta\,\hat{x}+(|\varepsilon|^2+|\mu\varepsilon-\sin^2\theta|-\sin^2\theta)\hat{z}\}, \quad (34a)$$

$$F^{(\text{surface})} = \frac{\varepsilon_o|E_o|^2\sin\theta\cos^2\theta}{|\sqrt{\mu\varepsilon-\sin^2\theta}+\varepsilon\cos\theta|^2}\{2\operatorname{Re}[(\varepsilon^*-1)\sqrt{\mu\varepsilon-\sin^2\theta}]\hat{x}-(|\varepsilon|^2-1)\sin\theta\,\hat{z}\}, \quad (34b)$$

$$F^{(\text{total})} = \frac{\varepsilon_o|E_o|^2\cos^2\theta}{|\sqrt{\mu\varepsilon-\sin^2\theta}+\varepsilon\cos\theta|^2}\{2\operatorname{Re}[\varepsilon^*\sqrt{\mu\varepsilon-\sin^2\theta}]\sin\theta\,\hat{x}+(|\varepsilon|^2\cos^2\theta+|\mu\varepsilon-\sin^2\theta|)\hat{z}\}. \quad (34c)$$

The total force is readily seen to be $\tfrac{1}{2}\varepsilon_o|E_o|^2\cos\theta[(1-|\rho_p|^2)\sin\theta\,\hat{x}+(1+|\rho_p|^2)\cos\theta\,\hat{z}]$, which is the combined rate of change of the incident and reflected momenta. (The extra $\cos\theta$ factor is the ratio of the incident beam's cross-sectional area to its footprint at the slab's surface.) These formulas are valid for all values of $\varepsilon$ and $\mu$, whether complex or real. For a transparent medium having $\mu\varepsilon\geq\sin^2\theta$, the forces may be written in terms of the refracted angle $\theta'$, where $\sin\theta=\sqrt{\mu\varepsilon}\sin\theta'$, and the $E$-field magnitude inside the medium, namely,

$$|E_t| = \sqrt{E_x^2+E_z^2} = 2\sqrt{\mu\varepsilon}\,E_o\cos\theta/(\sqrt{\mu\varepsilon-\sin^2\theta}+\varepsilon\cos\theta). \quad (35)$$

Thus the bulk, surface, and total forces of Eqs. (34) may be re-written as follows:

$$F^{(\text{bulk})} = \tfrac{1}{4}\varepsilon_o|E_t|^2\{2\sin\theta'\cos\theta'\,\hat{x}+[1+(\varepsilon/\mu)-2\sin^2\theta']\hat{z}\}, \quad (36a)$$

$$F^{(\text{surface})} = \tfrac{1}{4}\varepsilon_o|E_t|^2\{2(\varepsilon-1)\sin\theta'\cos\theta'\,\hat{x}-(\varepsilon^2-1)\sin^2\theta'\,\hat{z}\}, \quad (36b)$$

$$F^{(\text{total})} = \tfrac{1}{4}\varepsilon_o|E_t|^2\{2\varepsilon\sin\theta'\cos\theta'\,\hat{x}+[1+(\varepsilon/\mu)-(1+\varepsilon^2)\sin^2\theta']\hat{z}\}. \quad (36c)$$

At normal incidence, Eq. (36a) reduces to $F^{(\text{bulk})}=\tfrac{1}{4}\varepsilon_o|E_t|^2[1+(\varepsilon/\mu)]\hat{z}$, which is consistent with Eq. (10). However, when the same beam propagates at an angle $\theta'$ relative to the surface normal, its cross-sectional area shrinks to $\cos\theta'$, and its momentum flux becomes

$$F^{(\text{flux})} = \tfrac{1}{4}\varepsilon_o|E_t|^2[1+(\varepsilon/\mu)]\cos\theta'(\sin\theta'\,\hat{x}+\cos\theta'\,\hat{z}). \quad (37)$$

The difference between this and the bulk force of Eq. (36a) is

$$\Delta F = F^{(\text{bulk})} - F^{(\text{flux})} = \tfrac{1}{4}\varepsilon_o|E_t|^2[1-(\varepsilon/\mu)]\sin\theta'(\cos\theta'\,\hat{x}-\sin\theta'\,\hat{z}), \quad (38)$$

which is the (unbalanced) force on the beam's lower sidewall in accordance with Eq. (24). Note that the $\sin\theta'$ factor in Eq. (38) is the excess length of the lower sidewall in Fig. 5(a). A more detailed discussion of this point is given in Ref. [7].

If the $E$-field contribution to the Lorentz force is expressed as $(\boldsymbol{P}\cdot\boldsymbol{\nabla})\boldsymbol{E}$ instead of $-(\boldsymbol{\nabla}\cdot\boldsymbol{P})\boldsymbol{E}$, then the force density on the sidewalls will be given by Eq. (28) rather than Eq. (24). Also, the contribution of surface charges to the total force will differ from that given by Eq. (34b). However, the sum of all the forces on the slab turns out to be the same, irrespective of which formula is used to compute the forces. Listed below are expressions for the modified forces in the system of Fig. 5(a), when the $E$-field contribution to the Lorentz force is written as $(\boldsymbol{P}\cdot\boldsymbol{\nabla})\boldsymbol{E}$.

$$F^{(\text{sidewall})} = \tfrac{1}{4}\varepsilon_o|E_t|^2[2\varepsilon-(\varepsilon/\mu)-1]\sin\theta'(\cos\theta'\,\hat{x}-\sin\theta'\,\hat{z}), \quad (39a)$$

$$F^{(\text{surface})} = -\tfrac{1}{4}\varepsilon_o|E_t|^2(\varepsilon-1)^2\sin^2\theta'\,\hat{z}. \quad (39b)$$

It is easy to verify that the total force (per unit surface area) of Eq. (36c) is the sum total of three terms: (i) the momentum flow rate into the medium, given by Eq. (37); (ii) the force on the excess length of the lower sidewall in Eq. (39a); (iii) the surface force given by Eq. (39b).

The point of the entire discussion in the present (and also the following) section is that the rate of flow of momentum inside a transparent material is independent of whether the beam arrives at the surface at normal incidence, as in Fig. 2, or at oblique incidence, as in Fig. 5. Once the forces that act upon the triangular region beneath the entrance facet of the slab shown in Fig. 5 have been accounted for, the light that leaves this triangle will carry the same momentum flux into the slab as the light that enters the slab at normal incidence.

## 8. Oblique incidence on a magnetic slab – case of s-polarization

When the incident beam on the semi-infinite slab is s-polarized, as in Fig. 5(b), the various field components in the free space and inside the slab will be given by (subscripts $i$, $r$, $t$ refer to incident, reflected, and transmitted beams):

$$\boldsymbol{E}_i = E_o \hat{\boldsymbol{y}} \exp\{\mathrm{i}2\pi f[(x\sin\theta + z\cos\theta)/c - t]\}, \tag{40a}$$

$$\boldsymbol{H}_i = Z_o^{-1} E_o (-\cos\theta\, \hat{\boldsymbol{x}} + \sin\theta\, \hat{\boldsymbol{z}}) \exp\{\mathrm{i}2\pi f[(x\sin\theta + z\cos\theta)/c - t]\}, \tag{40b}$$

$$\boldsymbol{E}_r = -\rho_s E_o \hat{\boldsymbol{y}} \exp\{\mathrm{i}2\pi f[(x\sin\theta - z\cos\theta)/c - t]\}, \tag{40c}$$

$$\boldsymbol{H}_r = -Z_o^{-1} \rho_s E_o (\cos\theta\, \hat{\boldsymbol{x}} + \sin\theta\, \hat{\boldsymbol{z}}) \exp\{\mathrm{i}2\pi f[(x\sin\theta - z\cos\theta)/c - t]\}, \tag{40d}$$

$$\boldsymbol{E}_t = E_o(1-\rho_s)\hat{\boldsymbol{y}} \exp\{\mathrm{i}2\pi f[(x\sin\theta + z\sqrt{\mu\varepsilon - \sin^2\theta})/c - t]\}, \tag{40e}$$

$$\boldsymbol{H}_t = Z_o^{-1} E_o [-(1+\rho_s)\cos\theta\, \hat{\boldsymbol{x}} + \mu^{-1}(1-\rho_s)\sin\theta\, \hat{\boldsymbol{z}}] \exp\{\mathrm{i}2\pi f[(x\sin\theta + z\sqrt{\mu\varepsilon - \sin^2\theta})/c - t]\}. \tag{40f}$$

In the above equations, the Fresnel reflection coefficient $\rho_s$ for an s-polarized plane-wave is

$$\rho_s = (\sqrt{\mu\varepsilon - \sin^2\theta} - \mu\cos\theta)/(\sqrt{\mu\varepsilon - \sin^2\theta} + \mu\cos\theta). \tag{41}$$

The force on the bulk of the slab is exerted on bound currents $\boldsymbol{J}_b$ and on the magnetization $\boldsymbol{M}$. The $E$-field contribution to the Lorentz force, whether expressed as $-(\nabla\cdot\boldsymbol{P})\boldsymbol{E}$ or $(\boldsymbol{P}\cdot\nabla)\boldsymbol{E}$, turns out to be zero for s-light. However, both terms in Eq. (3a) will be needed to account for the force experienced by $\boldsymbol{M}$. Unlike the case of p-light discussed in Sec. 7, there are no bound electric charges (or electric dipoles) at the entrance facet of the slab. Due to the discontinuity of $H_z$ at the entrance facet, however, the magnetic dipoles, via the term $\mu_o(\boldsymbol{M}\cdot\nabla)\boldsymbol{H}$, contribute a surface force. The expression of the Lorentz force density for s-light, given in Eq. (29), must be time-averaged and integrated over the (infinite) thickness of the slab to yield the force per unit surface area as follows:

$$\boldsymbol{F}^{(\mathrm{bulk})} = \frac{\varepsilon_o|E_o|^2 \cos^2\theta\,\{2\mathrm{Re}[\mu^*\sqrt{\mu\varepsilon - \sin^2\theta}]\sin\theta\,\hat{\boldsymbol{x}} + [|\mu|^2 + |\mu\varepsilon - \sin^2\theta| + (1-2\mathrm{Re}(\mu))\sin^2\theta]\,\hat{\boldsymbol{z}}\}}{|\sqrt{\mu\varepsilon - \sin^2\theta} + \mu\cos\theta|^2}, \tag{42a}$$

$$\boldsymbol{F}^{(\mathrm{surface})} = \frac{\varepsilon_o|E_o|^2 [2\mathrm{Re}(\mu) - |\mu|^2 - 1]\sin^2\theta\cos^2\theta\,\hat{\boldsymbol{z}}}{|\sqrt{\mu\varepsilon - \sin^2\theta} + \mu\cos\theta|^2}, \tag{42b}$$

$$\boldsymbol{F}^{(\mathrm{total})} = \frac{\varepsilon_o|E_o|^2 \cos^2\theta\,\{2\mathrm{Re}[\mu^*\sqrt{\mu\varepsilon - \sin^2\theta}]\sin\theta\,\hat{\boldsymbol{x}} + (|\mu|^2\cos^2\theta + |\mu\varepsilon - \sin^2\theta|)\,\hat{\boldsymbol{z}}\}}{|\sqrt{\mu\varepsilon - \sin^2\theta} + \mu\cos\theta|^2}. \tag{42c}$$

This total force is readily seen to be $\tfrac{1}{2}\varepsilon_o|E_o|^2\cos\theta[(1-|\rho_s|^2)\sin\theta\,\hat{\boldsymbol{x}} + (1+|\rho_s|^2)\cos\theta\,\hat{\boldsymbol{z}}]$, which is directly related to the rates of arrival and departure of the incident and reflected

momenta. As before, the extra $\cos\theta$ factor is the ratio of the incident beam's cross-sectional area to the footprint of the beam at the slab's surface. These formulas are valid for all values of $\varepsilon$ and $\mu$, whether complex or real.

In the case of a transparent medium having $\mu\varepsilon \geq \sin^2\theta$, the total force (per unit surface area) given by Eq. (42c) may be written in terms of the $E$-field magnitude inside the medium, $|E_t|=|E_y|$, and the refracted angle $\theta'$, where $\sin\theta = \sqrt{\mu\varepsilon}\sin\theta'$, as follows:

$$\boldsymbol{F}^{(\text{total})} = \tfrac{1}{4}\varepsilon_0|E_t|^2\{2\varepsilon\sin\theta'\cos\theta'\hat{\boldsymbol{x}} + [1+(\varepsilon/\mu)\cos^2\theta' - (\varepsilon\mu)\sin^2\theta']\hat{\boldsymbol{z}}\}. \tag{43}$$

Furthermore, this force can be written as the sum of the following three components:

$$\boldsymbol{F}_1 = \tfrac{1}{4}\varepsilon_0|E_t|^2[1+(\varepsilon/\mu)]\cos\theta'(\sin\theta'\hat{\boldsymbol{x}} + \cos\theta'\hat{\boldsymbol{z}}), \tag{44a}$$

$$\boldsymbol{F}_2 = \tfrac{1}{4}\varepsilon_0|E_t|^2[2\varepsilon-(\varepsilon/\mu)-1]\sin\theta'(\cos\theta'\hat{\boldsymbol{x}} - \sin\theta'\hat{\boldsymbol{z}}), \tag{44b}$$

$$\boldsymbol{F}_3 = \tfrac{1}{4}\varepsilon_0|E_t|^2[2\varepsilon-\varepsilon\mu-(\varepsilon/\mu)]\sin^2\theta'\hat{\boldsymbol{z}}. \tag{44c}$$

In these equations, $\boldsymbol{F}_1$ is the rate of flow of momentum along the transmitted beam in accordance with Eq. (10); here the beam propagates at angle $\theta'$ and has cross-sectional area $\cos\theta'$, exactly as in the case of p-light discussed in the preceding section. In fact, $\boldsymbol{F}_1$ of Eq. (44a) is the same as $\boldsymbol{F}^{(\text{flux})}$ of Eq. (37). The second force, $\boldsymbol{F}_2$, is the excess force exerted on the beam's lower sidewall; see Eq. (31) and the corresponding discussion in Ref. [7]. As before, the $\sin\theta'$ factor in Eq. (44b) is the excess length of the lower sidewall in Fig. 5(b). The third force, $\boldsymbol{F}_3$, is the surface force of Eq. (42b); just as the force density $(\boldsymbol{P}\cdot\nabla)\boldsymbol{E}$ in the case of p-polarized light gave rise to $\boldsymbol{F}^{(\text{surface})}$ of Eq. (39b), so is $\mu_0(\boldsymbol{M}\cdot\nabla)\boldsymbol{H}$, in the present case, producing the force $\boldsymbol{F}_3$ at the surface. The only non-zero component of $(\boldsymbol{M}\cdot\nabla)\boldsymbol{H}$ associated with the discontinuities at the entrance facet is $M_z(\partial H_z/\partial z)$. Now, the $B_z$ continuity allows one to determine $M_z$ from the discontinuity of $H_z$, namely,

$$M_z(x,t) = H_z(x, z=0^-, t) - H_z(x, z=0^+, t) = Z_0^{-1}|E_t|(1-\mu^{-1})\sin\theta\exp\{i2\pi f[(x/c)\sin\theta - t]\}. \tag{45}$$

The magnetic dipoles oriented along $z$ at the entrance facet are acted upon by $\mu_0 H_z(x, z=0^+, t)$ at one pole and by $\tfrac{1}{2}\mu_0[H_z(x, z=0^-, t) + H_z(x, z=0^+, t)]$ at the other pole. The net $H$-field acting on this dipole layer is thus given by

$$\Delta H_z = H_z(x, z=0^+, t) - \tfrac{1}{2}[H_z(x, z=0^-, t) + H_z(x, z=0^+, t)] = -\tfrac{1}{2}M_z(x,t). \tag{46}$$

The time-averaged surface force per unit area, $\langle F_z^{(\text{surface})}\rangle = \tfrac{1}{2}\text{Re}(\mu_0 M_z \Delta H_z^*)$, may now be shown to be given by Eq. (44c). The bottom line is that the total force of Eq. (43), exerted on a transparent slab by an s-polarized beam at oblique incidence, is compatible with the rate of flow of momentum inside the slab as given by Eq. (10). The proof of this statement, however, required an analysis of the forces exerted on the medium within the triangular region immediately beneath the entrance facet of the slab of Fig. 5(b). In the above discussion, we showed that $\boldsymbol{F}_2$ and $\boldsymbol{F}_3$ of Eq. (44) are associated with the forces exerted on this triangular region, leaving the remaining term $\boldsymbol{F}_1$ to account for the momentum flux along the beam's propagation direction.

## 9. Concluding remarks

An important result of the present paper is that Eq. (3a), augmented by the additional term, $\partial(\boldsymbol{E}\times\boldsymbol{M}/c^2)/\partial t$, describes the force of the electromagnetic field on a material's magnetization density $\boldsymbol{M}$. To demonstrate the validity of this formula, we compared the total force exerted on a semi-infinite slab illuminated by a plane-wave (both at normal incidence and at oblique incidence) with the rates of flow of the incident and reflected momenta. Complete agreement between the two methods of calculation was obtained in every case. Along the way, we obtained expressions for (i) the momentum density of electromagnetic waves inside linear, isotropic, homogeneous materials, Eq. (11); (ii) the photon momentum inside transparent magnetic media, Eq. (12); and (iii) the lateral force experienced by a host medium at the sidewalls of a finite-diameter beam, Eqs. (24), (28) and (31). Also, relying on an "Einstein box" Gedanken experiment, we concluded that the "intrinsic" mechanical momentum density ½$E \times M/c^2$ of a magnetic medium arises from the interaction between $M$ and the electric component of the electromagnetic field.

In an isotropic medium whose polarization and magnetization densities are denoted by $P$ and $M$, the total force exerted by the electromagnetic field is the sum of the forces experienced by $P$ and $M$. The density of the $E$-field's force on bound electric charges can be written either as $-(\nabla \cdot P)E$ or as $(P \cdot \nabla)E$; the two formulations yield identical results for the total force (and total torque) on a given solid object, provided that the forces acting at the boundaries of the object are properly taken into account [12, 15]. When the force of the $E$-field on the bound electrical charges is written as $(P \cdot \nabla)E$, the complete expression of the total force density will be

$$F_{\text{total}}(x, y, z, t) = (P \cdot \nabla)E + (\partial P/\partial t) \times B + \mu_o M \times (\nabla \times H) + \mu_o (M \cdot \nabla)H + \partial(E \times M/c^2)/\partial t. \quad (47)$$

A similar formula, of course, could be written based on the alternative form, $-(\nabla \cdot P)E$. Now, using Maxwell's equations to combine its various terms, Eq. (47) can be streamlined into the following "generalized Lorentz law" for isotropic media:

$$F_{\text{total}}(x, y, z, t) = (P \cdot \nabla)E + \mu_o (M \cdot \nabla)H + \mu_o (\partial P/\partial t) \times H - \mu_o \varepsilon_o (\partial M/\partial t) \times E. \quad (48)$$

This generalized form of the Lorentz law has an aesthetically pleasing symmetry between the contributions of $P$ and $M$ to the overall force density. It also indicates that, as far as electromagnetic force is concerned, the relevant fields are $E$ and $H$ (rather than $B$ and $D$). This is noteworthy, considering that the same fields ($E$ and $H$) also appear in the expression of the Poynting vector $S$.

All the major results of the present paper can be derived directly from the generalized Lorentz law of Eq. (48). Momentum conservation among the incident, reflected and transmitted beams, which has been demonstrated throughout the paper for cases of normal and oblique incidence, is a significant piece of evidence in support of Eq. (48). Moreover, when computed in accordance with Eq. (48), the mechanical momentum imparted by an electromagnetic pulse to a transparent slab is found to be fully consistent with the requirements of the Einstein box Gedanken experiment outlined in Sec. 4. In other words, there will be no *hidden* momentum in the slab; a conclusion that lends further credence to Eq. (48).

Although justification for the form of Eq. (48) may be found in the theory of quantum electrodynamics, as has been attempted by Shockley [20], it is also possible to accept this generalized form of the Lorentz law in the same way that the original form of the law, or, for that matter, Maxwell's equations themselves, have been accepted, namely, as a law of nature.

We close by pointing out that Eq. (20), the expression of the total momentum density of a plane-wave in a host medium, is suggestive as to the nature of the momentum of light in vacuum. If the vacuum is assumed to have acquired polarization $P = \varepsilon_o E$ and magnetization $M = H$ in the presence of the $E$ and $H$ fields, then the optical momentum density of ½$E \times H/c^2$ in vacuum could be said to have arisen by equal contributions from this $P$ and $M$ in the form of ¼$\mu_o P \times H$ and ¼$E \times M/c^2$.

**Acknowledgements**

This work has been supported by the Air Force Office of Scientific Research (AFOSR) under contract number FA 9550−04−1−0213. The author is grateful to Ewan Wright, Khanh Kieu, and Rodney Loudon for helpful discussions.